\documentclass[twoside,a4paper]{article}

\usepackage[sc]{mathpazo} 
\usepackage[T1]{fontenc} 
\usepackage{microtype} 

\usepackage[pdftex]{graphicx}

\usepackage[hmarginratio=1:1,top=32mm,columnsep=20pt]{geometry} 
\usepackage{multicol} 
\usepackage{fancyhdr} 

\usepackage[hang, small,labelfont=bf,up,textfont=it,up]{caption} 
\usepackage{booktabs} 
\usepackage{float} 

\usepackage{lettrine} 
\usepackage{paralist} 

\usepackage{abstract} 

\usepackage[compact]{titlesec}

\titleformat{\section}{\Large\bfseries}{\thesection}{0.5em}{}
\titleformat{\subsection}{\large\bfseries}{\thesubsection}{0.4em}{}
\titleformat{\subsubsection}{\normalsize\bfseries}{\thesubsubsection}{0.3em}{}

\makeatletter
\newenvironment{tablehere}
  {\def\@captype{table}}
  {}

\makeatother

\title{\vspace{-15mm}\fontsize{24pt}{10pt}\selectfont\textbf{Resource management on a VM based computer cluster for scientific computing}} 

\author{
\large
\textsc{Stefano Stalio, Giuseppe Di Carlo, Sandra Parlati, Piero Spinnato}
\\[2mm] 
\normalsize Istituto Nazionale di Fisica Nucleare - Laboratori Nazionali del Gran Sasso \\ 
\normalsize \href{mailto:stefano.stalio@lngs.infn.it}{[stefano.stalio\footnote{corresponding author}, giuseppe.dicarlo, sandra.parlati, piero.spinnato]@lngs.infn.it} 
\vspace{-5mm}
}
\date{}

\begin{document}

\maketitle 

\thispagestyle{fancy} 


\begin{abstract}
\noindent
In the last ten years host virtualization has brought a revolution in
the way almost every activity related to information technology is
thought of and performed. The use of virtualization for HPC and HTC computing,
while eagerly desired, has probably been one of the last steps of this
revolution, the performance loss due to the hardware abstraction layer being
the cause that slowed down a process that has been much faster in other
fields.\\
Nowadays the widespread diffusion of virtualization and of new
virtualization techniques seem to have helped breaking this last
barrier and virtual host computing infrastructures for HPC and HTC are found in many data centers.\\
In this document the approach adopted at the INFN {\textquotedblleft}Laboratori Nazionali del
Gran Sasso{\textquotedblright} for providing computational resources via a virtual host based computing
facility is described. Particular evidence is given to the storage
layout, to the middleware architecture and to resource allocation
strategies, as these are issues for which a personalized solution was
adopted. Other aspects may be covered in the future within other documents.

\end{abstract}


\begin{multicols}{2} 
\section{Introduction}
The first tests with a virtual host (VM) based scientific computing environment were
performed at the {\textquotedblleft}Laboratori Nazionali del Gran
Sasso\footnote{http://www.lngs.infn.it}{\textquotedblright} (LNGS) around the second half
of year 2008. Starting from year 2010 the IT department staff has been
working at the development of a VM based computer cluster to be used by
the local experimental collaborations for their computational needs.
A constraint was set to only employ open source software tools for every 
aspect of the cluster management.\\ 
The aforementioned computing facility is in production at LNGS since July 2011
as part of U-LITE\footnote{http://ulite.lngs.infn.it} (Unified LNGS IT
Environment), a broader project that aims at creating an integrated IT
infrastructure that can be used to manage the whole life cycle of data produced by 
LNGS experiments:  storage and backup, long term archiving, distribution over the 
network, plus data analysis and detector simulation.\\ 
The U-LITE computing cluster must be able to dynamically provide
computing resources in the form of virtual hosts in response to user
requests that come in the form of jobs submitted to a batch queue
system. The same hardware infrastructure must be used by different
workgroups, each running its jobs on a customized software platform -
VMs cloned from a dedicated template - and sharing fairly and
efficiently a limited set of computing resources.\\
The U-LITE computing cluster was designed to meet the very specific
requirements of a restricted community of researchers, with no plan to 
be distributed outside LNGS, nevertheless we think that 
the approach we adopted can be of interest
for persons involved in host virtualization for HPC and HTC\footnote{HPC is, as 
defined by the European Grid Infrastructure, {\textquotedblleft}a computing paradigm that focuses on
the efficient execution of a large number of loosely-coupled tasks{\textquotedblright}, 
while HPC systems tend to focus on tightly coupled parallel jobs, and as such must execute within 
a particular site with low-latency interconnects.} computing.\\ 

\begin{figure*}
\begin{minipage}{15cm}
\includegraphics[width=15cm]{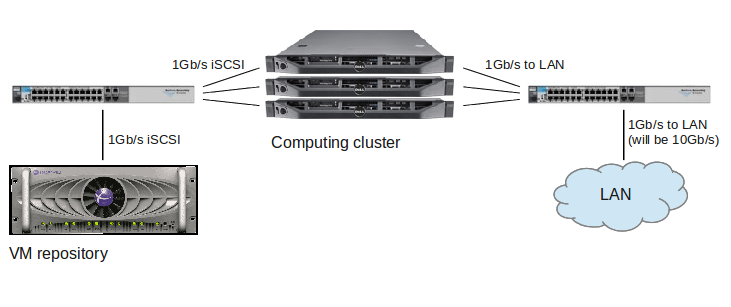}
\end{minipage}
\caption{The present network and storage
setup of the U-LITE computing cluster. Besides improving the uplink
towards the LNGS LAN to 10Gb/s, also the uplink towards the VM
repository should offer a higher throughput, particularly if the
computing power of the cluster should increase.}
\label{fig:storage layout}
\end{figure*}

\section{The cluster architecture}
The U-LITE computing cluster is made of a relatively small number of
heterogeneous multicore computers\footnote{12 servers for a grand
total of 216 CPU cores as of today} running the
Debian\footnote{http://www.debian.org} based Proxmox
VE\footnote{http://www.proxmox.com} virtualization platform and using the KVM hypervisor. 
The only function of the physical servers is to provide hardware resources for
the VMs they house.\\
The cluster computers intercommunicate on a dedicated LAN that has a
gateway towards the outside world and are connected via an iSCSI SAN to
a shared storage system that serves VM images (figure ~\ref{fig:storage layout}).\\
All VMs are clones of a limited number of templates, each template being
set up for a project or for a workgroup.

\subsection{The storage model}
The storage area that contains VM images is shared among all the cluster
computers, VM images are stored as LVM logical volumes and are available
to all the cluster computers as well. This means that the migration of
a VM from one server to another is always possible and, if it is
powered off, it only requires the few seconds needed to transfer the
VM configuration file from the source server to the destination.\\
This storage setup is very flexible and allows for very fast VM
provisioning but it is not the one that offers the best performances as
far as disk I/O is concerned. A few points must be taken into account,
though:

\begin{compactitem}
\item VM images are only used to load the OS and sometimes the
applications to be run. Apart from this the I/O activity towards the
system disk is very limited.
\item I/O is usually performed over network file systems (NFS and AFS).
The computing cluster LAN bandwidth and the bandwidth of the uplink
towards the LNGS LAN are critical for I/O performance.
\item A storage area that is local to the hosting server can be created 
and mounted on a VM at boot time and destroyed when the VM is powered off. 
This storage area is available as a relatively high performance, 
temporary data buffer. 
\end{compactitem}
One disadvantage of keeping all VM images within the same storage system is
that it can become a single point of failure. Using
a RAID storage with redundant controllers and power supplies
reduces the risk of failure of the VM repository. Different storage
architectures deploying higher availability standards might be
investigated.

\begin{figure*}[!ht]
\begin{minipage}{15cm}
\includegraphics[width=15cm]{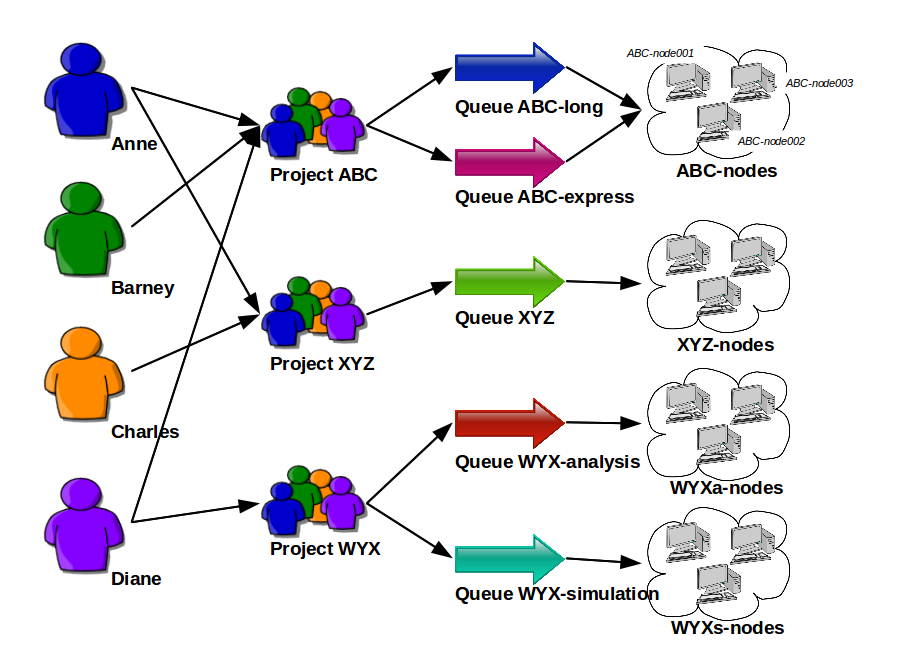}
\end{minipage}
\caption{Users, projects, queues and VM groups.}
\label{fig:relationships}
\end{figure*}

\section{Computing resources}
Each server belonging to the computing cluster provides hardware
resources to be allocated and used by VMs. These resources are:
\begin{compactitem}
\item RAM;
\item CPU cores;
\item local disk space for temporary data storage.
\end{compactitem}
A server usually makes all its CPU cores available for VM hosting as
well as a configurable portion of a local disk and almost all of its
RAM, reserving a relatively small amount of it (512MB or 1GB are
reasonable values) for processes running on the server OS itself.\\
VM slots are an extra resource type, not strictly related to the server hardware, 
that could be used to limit the number of VMs to be run on a single server. 
The VM slot number (which can be chosen arbitrarily) associated to each server
represents the maximum number of VMs it should host.\\
Each running VM allocates a configurable quantity of RAM, a configurable
number or CPU cores and optionally some local disk space. Besides that,
it occupies one VM slot on the hosting server. A VM can only be started
on a server if the latter has enough resources available among those
dedicated to VM hosting.\\
\section{Software architecture}
Job submission is performed through a traditional batch queue system
based on the Torque\footnote{http://www.adaptivecomputing.com} software
for job delivery and on the
Maui\footnote{http://www.adaptivecomputing.com} scheduler for a fair
and balanced sharing of resources.\\
A third software element, called CRM, was developed at LNGS. CRM
collects information from the physical servers, from the Proxmox VE
cluster management system, from the Torque server and from the Maui
scheduler and operates with the goal of providing the resources
requested by users via the batch system and releasing unused resources.
CRM acts on VMs in the following ways:

\begin{compactitem}
\item it makes running VMs available or unavailable to the batch queue
system by toggling their offline\footnote{From the Torque Administrator
Guide: \par {\textquotedblleft}A common task is to prevent jobs from
running on a particular node by marking it offline with \textit{pbsnodes -o
nodename}. Once a node has been marked offline, the scheduler will no
longer consider it available for new jobs. Simply use \textit{pbsnodes -c
nodename} when the node is returned to service.{\textquotedblright} }
flag;
\item it powers VMs on, optionally migrating them before start-up, or
off; 
\item it deletes VMs or creates new ones as clones of an existing
template, when necessary.
\end{compactitem}
The Torque, Maui and CRM server processes all run on the same, dedicated
host, that can be a VM as well. As this node is of critical importance for 
the correct operation of the whole computing cluster, a strong effort has 
been made to have a replica node that can be activated in a few seconds 
in case of a planned downtime and activates automatically thanks to an high
availability mechanism if the primary node fails. The takeover
procedure does not affect the computing cluster operation and is
transparent to users.

\subsection{The batch queue system}
Resource requests are triggered by users submitting
jobs to a batch queue system. Batch job submission is
performed using the \textit{qsub} command belonging to the Torque software
suite. Users do not necessarily need to know how hardware resources are
accessed, they may very well think to be working on a
traditional system where all computing nodes are real
computers.\\
The main reason for using different queues is to allow users select 
the VM type, and thus the software platform, to employ for a specific job. 
Each batch queue is
associated to a single VM type, a group of VMs that are clones of the
same template. Each VM type can be associated to one or more queues.
Usually each VM type and its associated queues
{\textquotedblleft}belong{\textquotedblright} to a single project or
working group, while a single project or working group may
{\textquotedblleft}own{\textquotedblright} one or more queues and the
associated VM types (figure ~\ref{fig:relationships}).
The use of VMs that can be automatically powered on or off gives each 
user or collaboration the chance of using all the cluster capabilities 
when needed and allows for quick, automatic reallocation of idle 
resources.\\

\subsubsection{The job scheduler and the virtualized computing
environment}
The Torque and Maui batch job submission tools were developed long
before host virtualization took over, so Torque and Maui were not
designed to deal with computing nodes that are not always on-line and
with a system that is able to power computing nodes on or off in order
to optimize resource usage.\\
Also, Torque and Maui were born in an era
where computing resources were scarce and precious, so they were conceived
with the idea in mind that jobs must never exceed the maximum CPU time
allowed for the queue they run on and that users have to know how long
their jobs need to run and submit them to the appropriate queue. 
Today this scenario is obsolete: jobs must be able to run for as much
time as they need to complete successfully and users do not have to
worry about the resources their jobs need. Queues are used for
selecting the software platform (VM group) a job must run on or in
order to control the priority a computing job must be assigned.\\
Maui is highly configurable and is capable of sophisticated scheduling
decisions, particularly in environments where concurrent resource
requests must be evenly balanced. Its behavior though can be altered
in a situation where only a part of the computing resources seem
to be available for job submission.\\
In order to have Maui correctly perform resource sharing in such an environment, 
the status of the computing nodes must somehow be hidden to the scheduler, 
either by giving to the scheduler the illusion of having {\textquotedblleft}infinite{\textquotedblright}
resources at its disposal or by disabling all unused resources, checking how the
scheduler would prioritize queued jobs, and provisioning VMs for the
batch system in the same order.\\
The solution that was adopted at LNGS is an approximation of the latter.

\subsection{The Cluster Resource Manager}
As briefly mentioned before, CRM administers hardware resources by
powering VMs on or off and migrating them if needed before starting them up.
Hot migration of powered on VMs towards faster servers has been
implemented but it has not been activated in the production environment
for the following reasons:

\begin{compactitem}
\item the CPU speed span among the cluster computers is not so high as to
justify the activation of such a feature;
\item it is quite difficult to define a hot migration algorithm that is 
able to safely and efficiently optimize resource usage.
\item hot migration of powered-on VMs is not considered reliable enough
to be activated in a production environment (e.g. hot migration of active 
nodes may fail if the servers software versions are not aligned);
\end{compactitem}
As mentioned earlier, CRM does not only turn VMs on or off. A mechanism 
has been implemented that reduces the computing overhead due to boot and shutdown
procedures: VMs can be made unavailable/available to the batch queue
software by just setting/clearing their offline flag on Torque with
no need of power them off or on.\\
A very important task CRM is in charge of is making sure that the
resources shared by each physical server are not overbooked. This means
that the total amount of RAM, CPU cores and disk space allocated by the
VMs on a server must never exceed the resources that the server can
provide. It is particularly important that a portion of the server RAM is
always available for the processes running on its OS, otherwise memory
swapping may occur, making the server and its VMs unusable or even
causing processes on the server (VMs are in fact processes on a server)
die for lack of RAM space.\\
The computing cluster is not strictly reserved for VMs dedicated to
batch job processing. CRM will take into account resources
allocated by VMs not belonging to the batch system but will not modify
their status. The same computing environment can thus be used both for batch
job processing and for the delivery of other network services.

\begin{figure*}[!ht]
\begin{minipage}{13cm}
\includegraphics[width=13cm]{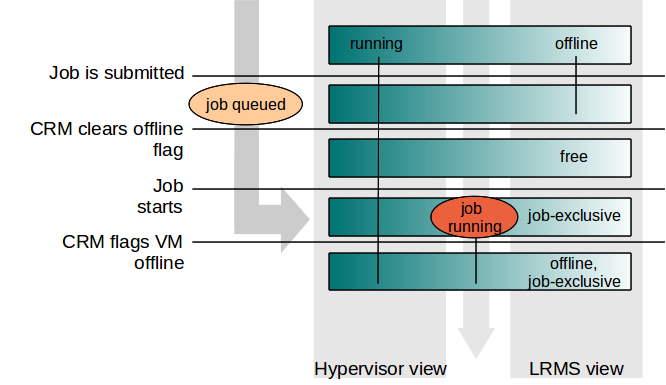}
\end{minipage}
\caption{When a job is submitted and is eligible to run on a powered on VM, the offline flag is cleared and is set again as
soon as the job starts. This technique reduces the job queue
time without influencing the scheduler operation. LRMS (Local Resource Management System) is the batch queue system.}
\label{fig:toggle offline before}
\end{figure*}

\begin{figure*}[!ht]
\begin{minipage}{13cm}
\includegraphics[width=13cm]{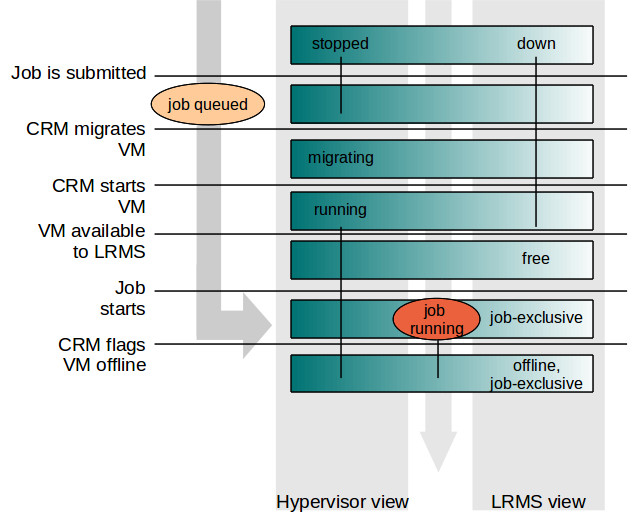}
\end{minipage}
\caption{How a VM is powered on in order to make it available for a job to run.}
\label{fig:power on}
\end{figure*}

\section{Resource management}
\subsection{Providing resources}
Whenever a job is submitted it is given by Maui a
priority which depends on the scheduler setup. At every scheduler
iteration (typically once or twice a minute), queued job priorities are
updated. The calculation of job priorities can take into account many
parameters and is highly configurable by administrators\footnote{See
Maui documentation at
http://www.adaptivecomputing.com/resources/docs/maui/mauiadmin.php}.\\
At every iteration\footnote{Reasonably the CRM execution period should
be the same as Maui. It would be useful to have Maui trigger a CRM
iteration at the end of its own.} CRM obtains the list of queued jobs, 
ordered by priority, from the job scheduler and takes care of providing 
resources according to the list order.\\
In order to optimize the response time to job requests, for each job it
processes, CRM first checks if there is a suitable running VM and, 
if it finds one, it clears the VM offline flag thus making it ready 
to run jobs (figure ~\ref{fig:toggle offline before}). 
When the system is idle, or lightly loaded, a job queue time is around 30 to 35 
seconds if CRM only needs to toggle the VM offline flag.\\
If CRM does not find a running VM for the job it is processing,
it starts one that is powered off (figure ~\ref{fig:power on}), given that there are enough
available hardware resources on one of the physical servers. 
CRM optionally migrates each VM, before starting it, towards the fastest server, 
in terms of CPU frequency, that has enough resources to host it.
 
\begin{compactitem}
\item The total sum of CPU cores allocated by the VMs should never
exceed the total number of CPU cores mounted on the server they run on.
The computing resources required by the guest OS are considered
negligible. If, due to a clumsy manual intervention, the number of CPU
cores allocated by the VMs should exceed the total number or CPU cores
of the server the only consequence will be a non-optimal resource
usage.
\item The total quantity of RAM allocated by VMs should always be less
than the total quantity of physical RAM mounted on the server. It is
very important that a reasonable quantity of RAM is reserved for the
processes running on the guest server OS, otherwise it may start using 
the swap area on disk and start performing unacceptably slow together with
the VMs it hosts .
\item The total quantity of disk space, if any, allocated by VMs should
never exceed the size of the partition dedicated to this purpose. The
system partition should not be used for disk images.
\end{compactitem}
In the case CRM needs to power on a VM before a queued job
can be started, and in case of low cluster load, the job queue time is
around 90 to 120 seconds.\\
If there is no suitable VM on the system, there are enough available
hardware resources on the computing cluster, and there is enough disk
space on the storage system dedicated to hosting VM images, CRM can
create one from an existing template. In this case the job queue time
depends on the size of the VM image and may range from 5 to 15 minutes.\\
If the storage area that holds VM images is large enough compared to
the cluster size, the creation and/or destruction of a VM is expected
to be a rare event.

\begin{figure*}[!ht]
\begin{minipage}{13cm}
\includegraphics[width=13cm]{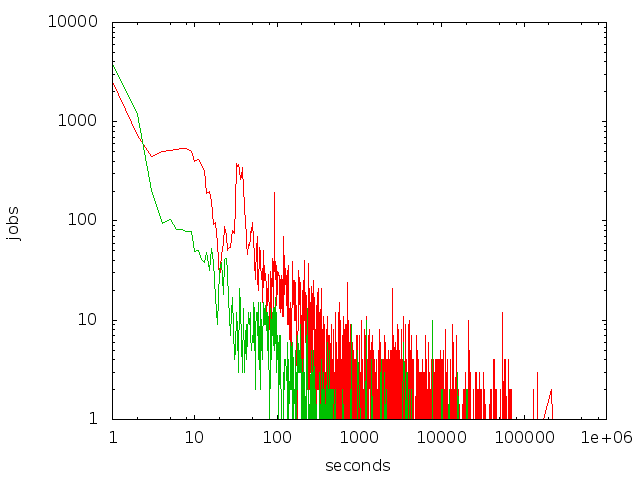}
\end{minipage}
\caption{The job queue time distribution
in the last year for physical servers (green) and VMs (red).}
\label{fig:queue time}
\end{figure*}

\begin{tablehere}
\caption{Approximate job queue time in case of low cluster load (figure ~\ref{fig:queue time}) 
}
\centering
\begin{tabular}{|p{3cm}|p{3cm}|} \hline
VM status & Queue time \\\hline 
free & lt 10 s \\
offline & 30/40 s \\ 
powered off & 60/90 \\
to be instantiated & 300/600 s \\\hline
\end{tabular}
\end{tablehere}

\subsection{Releasing resources}

\begin{figure*}[t]
\begin{minipage}{13cm}
\includegraphics[width=13cm]{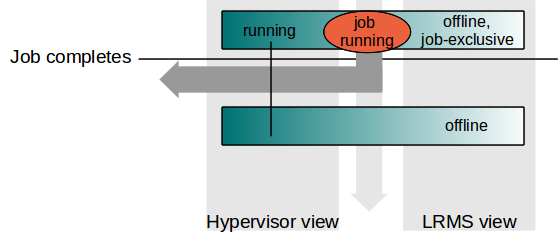}
\end{minipage}
\caption{If there are enough idle
resources VMs are only flagged offline instead of being powered off.}
\label{fig:toggle offline after}
\end{figure*}

\begin{figure*}[!ht]
\begin{minipage}{13cm}
\includegraphics[width=13cm]{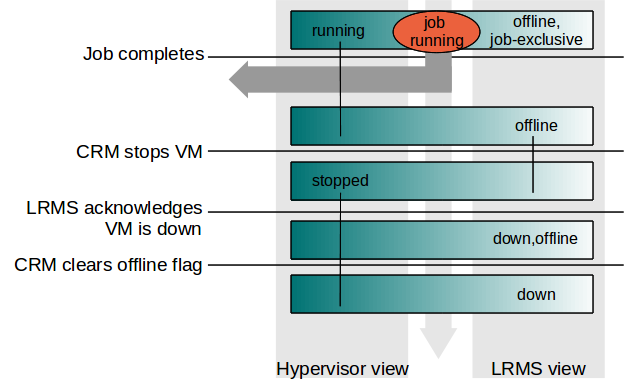}
\end{minipage}
\caption{How a VM is powered off at job completion.}
\label{fig:power off}
\end{figure*}

Whenever a job starts on a VM, CRM marks the latter offline for the
batch job scheduler. This means that the VM is not available for new
jobs to run. When the job ends the VM remains powered on but with the
offline flag set (figure ~\ref{fig:toggle offline after}).\\
If there are queued jobs requiring a different VM type, idle VMs may be
powered off (figure ~\ref{fig:power off}) in order to release resources (CPU cores, RAM, local disk
space) and may be destroyed if there is need of disk space for new VMs
on the dedicated storage area.\\
If there are unused VMs running, a few of them can be powered off or
destroyed in order to free a configurable amount of RAM, number of CPU
cores and disk space with the goal of improving the system
response time when jobs requiring nodes that are powered off will be
queued.\\
CRM also employs a mechanism by which it powers off the first VM
that becomes idle on a server that, for any reason, has been
overbooked. 

\section{Extensions to the model}
The mechanism for administering hardware resources described in the
previous paragraphs has been designed so that it does not influence the
job scheduling process. Resource sharing among users or
working groups is always in full control of the scheduler.\\ 
The model just described has been extended to cope with different needs
and to optimize resource usage.

\begin{compactitem}
\item Part of the computing nodes, be they VMs or real computers (with
no virtualization support activated) can be configured to never have their offline 
flag set, never be powered off, and always be ready to accept jobs. This
allows working groups to have a set of resources reserved for their
needs and always ready for use. It also allows working groups to
integrate those hardware resources that for some reason cannot be
used for host virtualization into the common computing environment.
\item The computing capabilities of each VM can be extended by
increasing the number of CPU cores (and RAM size accordingly) assigned
to it. This feature has quite important implications that are better
described in the next section.
\end{compactitem}

\subsection{Multicore VMs}
Enabling multicore VMs has some evident advantages:
\begin{compactitem}
\item multiple jobs can run on the same VM concurrently;
\item multi processor and parallel jobs are supported;
\item less disk space must be allocated to VM images on the dedicated
storage area;
\item the CPU, memory and network overhead due to running operating
systems and system services is reduced;
\end{compactitem}
On the other hand, the fact that a VM can run multiple jobs concurrently 
has a negative impact on the granularity of the resource sharing mechanism.
Suppose that the VMs associated to a certain queue are configured to start with
4 CPU cores and that one of them is started in order to provide resources
for one single processor job. Should CRM flag the node offline after
the first job starts running, three CPU cores would be unavailable to
the batch system and resource utilization would be far from optimal.
Presently CRM does not flag VMs offline after the start 
of the first job and the idle CPU cores are allocated to the
queue or group of queues the hosting VM is associated to. The resource 
sharing system may become unbalanced because these computing resources are 
seen by the batch scheduler as immediately available and the workgroup
{\textquotedblleft}owning{\textquotedblright} them is privileged in the short term.\\
The Maui scheduler can be
configured so to balance resource usage in the medium or long term,
but this short term unbalance might still be a problem in periods of
high workload. In the worst case, if one workgroup keeps on submitting
jobs to its queues not allowing the associated VMs to be powered off,
jobs belonging to other collaborations may never get the slots they
need.\\
One solution would be to only allow jobs to be submitted to a VM for a
fixed time window after power on and to set it offline after the end of 
this time window, regardless of the number of jobs it is running or has run. 
This stratagem would avoid starvation of jobs belonging to other queues 
but would not optimize resource usage. The code implementing 
this feature has been written but has not been activated on CRM as serious 
job starvation issues never showed up yet. 

\section{Conclusions and future work}
The development of the U-LITE computing cluster started in 2010 and
even though new features and functionalities are still being added, 
it is today a productive system and the official scientific computing 
facility at LNGS. Its strengths are the exclusive use of open source software, the 
fast provisioning of computing resources and the 
transparency from the user point of view.\\
The hardware and software architecture it is based on makes the
computing facility best fit for small to medium computing environments.
No critical scalability issues should show up if the cluster was to
grow by one order of magnitude in terms of computing capability, although some minor
infrastructural and software adjustments might be required, while
no prediction has been made on how it would behave and what changes would be required
if it was to grow further.\\
A lot of room remains for new features and developments. Items to work
on in the future, given the availability of human resources, include:

\begin{compactitem}
\item better definition of resource sharing policies and definition of a SLA with experimental collaborations; 
\item creation of a web portal for system monitoring and resource accounting, including detailed self job history for authenticated users;
\item run time customization of resources in accordance to user
requirements (e.g. a user wants to submit a job to a node with 2 computing cores, 
4 GBytes of RAM and 20 GBytes of temporary disk space and a dedicated VM is started accordingly);
\item automated customization of resources according to the job queue (e.g. if 6 jobs are queued for the same VM
type, start a VM with 6 CPU cores and adequate RAM size);
\item better VM placement algorithm;
\item investigation of VM checkpointing for improved resource balancing;
\item interfacing with standard cloud computing tools for better geographic interoperability.
\end{compactitem}

\end{multicols}
\end{document}